\documentclass[aps,prb,twocolumn,showpacs]{revtex4-2}

\usepackage{graphicx}
\usepackage{calc}
\usepackage{bm}
\usepackage{color}

\begin{document}

\title{Coalescence of Andreev bound states on the surface of a chiral topological semimetal}

\author{V.D.~Esin}
\author{Yu.S.~Barash}
\author{A.V.~Timonina}
\author{N.N.~Kolesnikov}
\author{E.V.~Deviatov}
\affiliation{Institute of Solid State Physics of the Russian Academy of Sciences, Chernogolovka, Moscow District, 
2 Academician Ossipyan str., 142432 Russia}

\date{\today}

\begin{abstract}
We experimentally investigate the magnetic field dependence of Andreev transport through a region of proximity-induced 
superconductivity in CoSi topological chiral semimetal. With increasing parallel to the CoSi surface magnetic field, 
the sharp subgap peaks, associated with Andreev bound states, move together to nearly-zero bias position, while there is
only monotonic peaks suppression for normal to the surface fields. The zero-bias $dV/dI$ resistance value is perfectly 
stable with changing the in-plane magnetic field. As the effects are qualitatively similar for In and Nb superconducting
leads, they reflect the properties of a proximized CoSi surface.  The Andreev states 
coalescence and stability of the zero-bias $dV/dI$ value with increasing in-plane magnetic field are interpreted as the 
joined effect of the strong spin-orbit coupling and the Zeeman interaction, known for proximized semiconductor nanowires. We associate the observed magnetic field anisotropy with the  recently predicted in-plane polarized spin texture of the Fermi arcs surface states. 
\end{abstract}

\maketitle


Recently, chiral topological semimetals have been predicted~\cite{bernevig,zhang} as natural generalization of Weyl 
semimetals~\cite{armitage}. They are characterized by simultaneously broken mirror and inversion symmetries and non-zero
Chern numbers. Chiral topological semimetals host new types of massless fermions with a large topological charge, which 
lead to numerous exotic physical phenomena like unusual magnetotransport~\cite{magnet}, lattice dynamics~\cite{lattice},
and a quantized response to circularly polarized light~\cite{photo}.

In topological semimetals, the nontrivial topology results in  extensive Fermi arcs connecting projections of  bulk 
excitations on the side surface~\cite{armitage}. In a chiral topological semimetal there is only one pair of chiral 
nodes of opposite Chern numbers with large separation in momentum space. This leads  to extremely long surface Fermi 
arcs~\cite{long}, in sharp contrast to Weyl semimetals, which have multiple pairs of Weyl nodes with small 
separation~\cite{armitage}.

Chiral topological semimetals can be realized, in particular, in a family of transition metal silicides with a chiral 
crystal structure, including CoSi, RhSi, RhGe, and CoGe single crystals, where CoSi is the mostly investigated material.
Bulk band structure and extremely long Fermi arcs have been  confirmed by angle-resolved photoemission spectroscopy 
(ARPES)~\cite{long,cosi1,cosi2,maxChern}.

In proximity to a superconductor, topological materials exhibit non-trivial physics that can in various cases result
in topological superconductivity and existence of Majorana modes ~\cite{review1,review2,LiXu2019,volovik}. This concerns not 
only the topological insulators~\cite{zhang1,kane,zhang2,Fu}, but also Weyl semimetals, where the proximity
was predicted to produce specular Andreev reflection~\cite{spec} (similar to the graphene case~\cite{been1,been2}), 
superconducting correlations and Majorana modes in the Fermi arcs~\cite{Franz2016,arcsc} as well as various 
superconducting pairings decaying in the depth of the sample~\cite{dutta2020}. Topological surface states are 
responsible for Josephson current for long superconductor-semimetal-superconductor junctions~\cite{inwte1,inwte2,incosns}, 
and Tomasch oscillations within the region of proximity-induced superconductivity~\cite{nbwte}.

A proximity-induced superconductivity in chiral topological semimetals with multifold fermions, such as CoSi, has been 
studied until now neither experimentally nor theoretically. Although, a superconducting state allow the existence of topological 
superconductivity with surface Majorana fermions~\cite{supercond} in a doped chiral  semimetal interfaced with the undoped one.

Here, we investigate the magnetic field dependence of Andreev transport through a region of proximity-induced superconductivity 
in CoSi chiral topological semimetal. We observe sharp subgap peaks, which are usually
ascribed to Andreev bound state (ABS) positions. Evolution of these peaks depends on the magnetic field orientation: they are 
moving together to nearly-zero bias position for parallel to the CoSi flake surface magnetic fields, while there is only
monotonic peaks suppression in normal magnetic fields. Also, zero-bias $dV/dI$ resistance value is perfectly stable in 
parallel magnetic field. These effects are qualitatively similar for In and Nb superconducting leads, so they reflect 
properties of a proximized CoSi surface. 

The behavior of the peaks with increasing in-plane magnetic field can be interpreted as ABSs coalescence due to the 
joined effect of spin-orbit coupling (SOC) and Zeeman interaction. The effect is known for proximized semiconductor
nanowires \cite{sarma}. The observed magnetic field anisotropy can be associated with the Zeeman interaction of the 
Fermi arcs states on (001) surface in CoSi, which have recently been predicted to be in-plane spin polarized 
\cite{Burkovetal2018}.


\begin{figure}
\includegraphics[width=1\columnwidth]{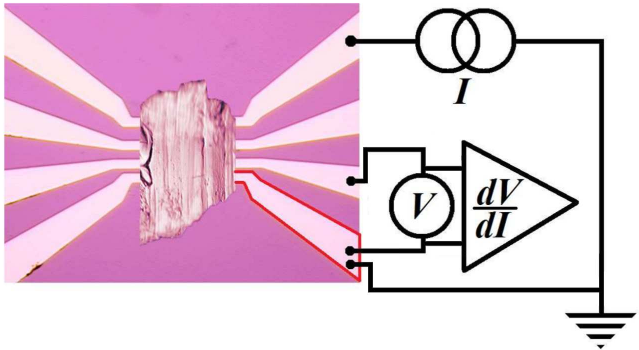}
\caption{(Color online) Top-view image of the sample with schematic diagram of electrical connections. A small (about 
100~$\mu\mbox{m}$ wide and 1~$\mu\mbox{m}$ thick) single-crystal CoSi ﬂake is placed on the pre-defined Nb leads 
pattern. We investigate transport through 5~$\mu$m long  CoSi region between two 10~$\mu$m wide superconducting leads: one lead is 
grounded, two other leads are employed to apply current $I$ and measure voltage drop $V$, respectively. To obtain 
$dV/dI(I)$ characteristics,  the dc current  is additionally modulated by a low ac component. All the wire resistances 
are excluded, which is necessary for low-impedance samples.
}
\label{sample}
\end{figure}

The initial CoSi material was synthesized from cobalt and silicon powders by 10$^\circ$~C/h heating in evacuated silica
ampoules up to 950$^\circ$~C. The ampoules were held at this temperature for two weeks and then cooled down to room 
temperature at 6$^\circ$~C/h rate. The obtaibed material was identified  as CoSi with some traces of SiO$_2$ by X-ray 
analysis. Afterward, CoSi single crystals are grown from this initial load by iodine transport in evacuated silica 
ampoules at 1000$^\circ$. X-ray diffractometry demonstrates cubic structure of the crystals, also,  X-ray spectral 
analysis confirms equiatomic ratio of Co and Si in the composition, without any SiO$_2$ traces.

To investigate transport through a region of proximity-induced superconductivity in CoSi topological semimetal, we use some modification of standard thin-flake sample preparation technique~\cite{inwte1,inwte2,incosns,nbwte,cdas}. Topological semimetals are essentially three-dimensional crystals~\cite{armitage}, so one has to use thick CoSi flakes.  Small flakes can be easily obtained from the initial CoSi single crystal by a mechanical cleaving 
method~\cite{incosns,cdas}. We determine the flake surface as (001) one from standard magnetoresistance  measurements~\cite{magres}. Van der Waals forces are too weak to hold a  1~$\mu\mbox{m}$ thick flake on the superconducting (niobium or indium) contacts, thus, it is pressed  slightly with another oxidized silicon substrate, see  Fig.~\ref{sample}. The contacts pattern consists of 10~$\mu$m wide superconducting leads, which  are defined by lift-off technique on the insulating SiO$_2$ substrate after magnetron sputtering of 150~nm Nb or thermal evaporation of 100~nm In.

This procedure provides transparent normal-superconductor (NS) junctions, stable in different cooling cycles~\cite{inwte1,inwte2,incosns,nbwte,cdas}. In present experiment, both Nb-CoSi-Nb and  In-CoSi-In junctions are characterized by low (0.6~Ohm and 1.5~Ohm, respectively) junction resistances, which indicate broad ($\approx 10\times10 \mu$m$^2$ area) planar junctions with high junction's transparency. For our sample  preparation technique, Andreev spectroscopy has been reliably demonstrated in Refs.~\cite{incosns,nbwte,cdas}. Moreover, direct experimental comparison of Andreev and thermal regimes of transport can be found in Ref.~\cite{cdas}.

We investigate transport through 5~$\mu$m long CoSi region between two superconducting leads (Nb-CoSi-Nb or In-CoSi-In junctions), the
connection scheme is depicted in Fig.~\ref{sample}: one lead is grounded, two other leads are employed to apply current
$I$ and measure voltage $V$, respectively.  To obtain $dV/dI(I)$ characteristics,  the dc current (within $\pm$0.3~mA 
range)  is additionally modulated by a low ac component  (~5~$\mu\mbox{A}$, $f=7.7$~kHz). We measure both dc (V) and ac 
(which is proportional to $dV/dI(I)$) voltage components with a dc voltmeter and a lock-in, respectively.   Due to the superconducting leads,  all the wire resistances are excluded  in Fig.~\ref{sample}, which is necessary for low-impedance samples. The measurements are performed in the temperature interval 1.2 K – 4.2 K for two different magnetic field orientations in several cooling cycles.

\begin{figure}
\includegraphics[width=\columnwidth]{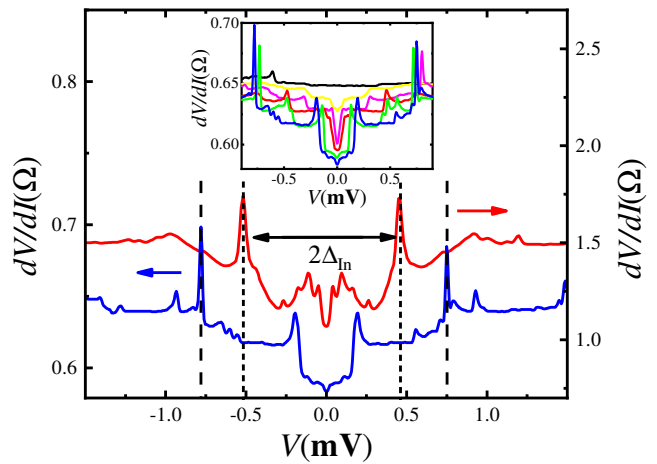}
\caption{(Color online) Typical examples of low-temperature $dV/dI(V)$ characteristics for Nb-CoSi-Nb (blue curve) and  
In-CoSi-In (red curve) junctions. The curves are qualitatively similar, superconducting gap positions are depicted by 
dashed lines.  In addition, different subgap $dV/dI(V)$ features  are known for finite-size NS junctions: while shallow oscillations  
originate from Tomasch and MacMillan-Rowell geometrical resonances, sharp subgap peaks are usually associated with Andreev bound states~\cite{mzm2,ABS}. Inset demonstrates $dV/dI(V)$  temperature dependence for Nb-CoSi-Nb junction, the curves are obrained at 1.4~K, 2.4~K, 3.6~K, 4.2~K, 5.0~K, 6.9~K, respectively. The flat $dV/dI(V)$ curve above 7~K well corresponds to the gap estimation $\Delta_{Nb}\sim$0.75~meV. 
}
\label{TempDepend}
\end{figure}


\begin{figure}
\includegraphics[width=0.8\columnwidth]{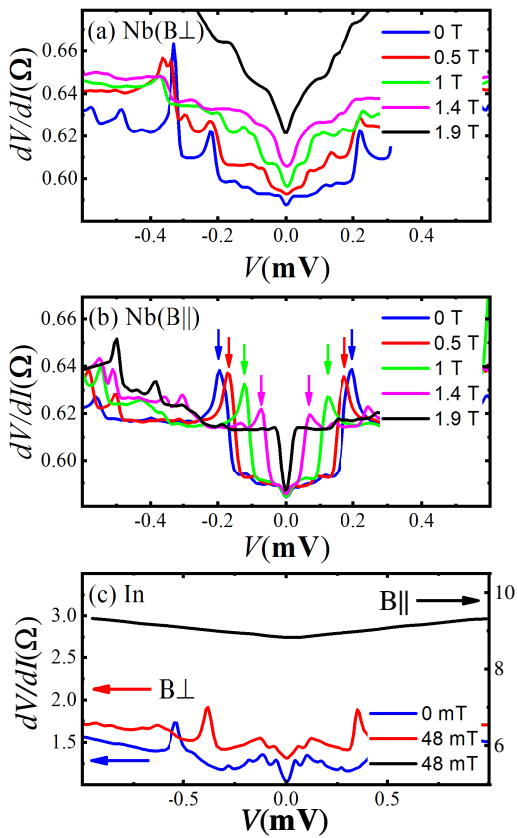}
\caption{(Color online) Evolution of $dV/dI(I)$ subgap region for two different orientations of magnetic field. (a) 
Nb-CoSi-Nb junction,  the field is oriented normally to the flake's plane. All subgap features are gradually suppressed,
$dV/dI$ level is increasing in magnetic field. (b) Nb-CoSi-Nb junction in parallel magnetic field. The zero-bias value 
$dV/dI(I=0)$ is stable, ABS peaks (arrows)  monotonicly come to nearly-zero bias. (c) Qualitatively similar behavior for 
In-CoSi-In junction: The zero-bias value $dV/dI(I=0)$  is very sensitive to normal magnetic field 48~mT, while the 
dependence is weak for the in-plane magnetic field of the same value.
}
\label{mag}
\end{figure}

 Fig.~\ref{TempDepend} demonstrates typical examples of low-temperature $dV/dI(V)$ characteristics for Nb-CoSi-Nb (blue 
curve) and  In-CoSi-In (red curve) junctions.   The curves are qualitatively similar, superconducting gap position is 
well defined, there are pronounced subgap features for junctions of both types.  
 
For the In-CoSi-In junction, the gap value well corresponds to the known~\cite{indium} one $\Delta_{In}\sim$0.5~meV, as 
depicted by the dashed line. Also,  all the features are suppressed above 3.5 K, which is known
critical temperature for indium~\cite{indium}. For Nb-CoSi-Nb junction,  the gap is diminished $\Delta_{Nb}
\sim$0.75~meV  as compared with the known bulk Nb value due to the non-perfect niobium film quality. This gap estimation is also 
supported by the temperature dependence in the inset to Fig.~\ref{TempDepend}: the $dV/dI(V)$ curve is flat above 7~K 
for the Nb-CoSi-Nb junction,  which is below the bulk 9~K value.  Observation of  well defined  superconducting gap is a direct confirmation of   Andreev 
regime~\cite{andreev,tinkham} of transport for  both type junctions. This identification of the superconducting gap is similar to one in Ref.~\cite{nbcosns}, where it is confirmed also by standard temperature and magnetic field  dependencies.
 
In the Andreev regime, different subgap $dV/dI(V)$ features, which can be seen in Fig.~\ref{TempDepend}, are known for finite-size 
junctions~\cite{ABS}. The pronounced wide central structure in $dV/dI$  reflects the proximity-induced 
gap~\cite{ABS,heslinga}, e.g. in the topological surface state~\cite{klapwijk17,ingasb}. Shallow oscillations 
originate~\cite{osbite}  from Tomasch~\cite{tomasch1,tomasch2} and MacMillan-Rowell~\cite{mcmillan1,mcmillan2} 
geometrical resonances or multiple Andreev reflection\cite{kuz1,kuz2}. In contrast, sharp subgap peaks are usually associated with Andreev bound states~\cite{mzm2,ABS}.

It is important, that these features (superconducting gap, oscillations, ABSs) can appear either as $dI/dV$ conductance peaks~\cite{mzm2} or $dV/dI$ resistance peaks~\cite{ABS}, depending on the experimental configuration. In particular, there are several type of carriers in topological materials (at least, bulk carriers and topological surface states)~\cite{klapwijk17}. For example, Ref.~\cite{ingasb} shows switching from $dI/dV$ conductance peaks to $dV/dI$ resistance peaks  for equally prepared In--InAs/GaSb interfaces, while InAs/GaSb is of normal or topological spectrum.  This difference originates from the fact, that charge transport does not directly reflects density of states. The main result of Ref.~\cite{klapwijk17} is that  scattering at different NS interfaces is responsible for the experimental $dV/dI(V)$ curves. The proximity-induced superconductivity can be expected within the conductive surface state  near the superconducting lead~\protect\cite{adroguer}. One NS interface is responsible for Andreev reflection at  biases below the induced gap, while above this value the NS interface with bulk superconductor governs the reflection process. Because of different single-particle transparencies of the interfaces, $dV/dI(V)$ contains~\protect\cite{klapwijk17} additional structures at low biases as $dI/dV$ conductance peaks~\cite{mzm2} or $dV/dI$ resistance peaks~\cite{ABS}, depending on the transparencies combination.

\begin{center}
\begin{figure}
\includegraphics[width=\columnwidth]{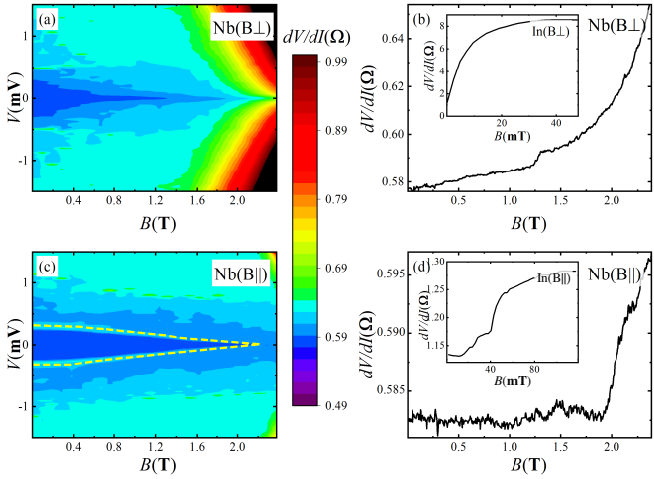}
\caption{
(Color online) (a) Detailed evolution of $dV/dI(V)$ level in normal magnetic field for Nb-CoSi-Nb junction. No special 
traces can be observed for ABS. (b) $dV/dI(I=0)$ level is monotonicly increasing in normal magnetic field 
scan for the Nb-CoSi-Nb junction (main field) and for the In-CoSi-In one (inset). (c) Subgap ABS peaks  monotonicly come to 
nearly-zero position in parallel magnetic field,  they are  coalescing together at approximately 2~T, as depicted by yellow dashed lines. (d) Zero-bias level 
$dV/dI(I=0)$  is stable in parallel magnetic field below 2~T for the Nb-CoSi-Nb junction (main field) and below 40~mT for the
In-CoSi-In one (inset). 
}
\label{colorplot}
\end{figure}
\end{center}

Our main experimental result is the difference in the ABS  evolution for two different orientations
of magnetic field, as it is demonstrated in Figs.~\ref{mag}, and~\ref{colorplot}. We trace $dV/dI$ resistance peaks as the ABS positions, following Ref.~\cite{ABS} due to the similar experimental setup.

For Nb-CoSi-Nb junction, central region of $dV/dI(I)$ curves is shown in  Fig.~\ref{mag} (a) and (b) for different 
magnetic fields. If the field is oriented normally to the flake's plane, all subgap features are gradually suppressed in
Fig.~\ref{mag} (a). $dV/dI$ level is monotonicly increasing, no special traces can be observed for ABS resonances, as 
it is shown by colormap in  Fig.~\ref{colorplot} (a) and by the $dV/dI(I=0)$ magnetic field scan in Fig.~\ref{colorplot}
(b). This behavior is usual for the  superconductivity suppression in magnetic field~\cite{tinkham}.

In contrast, the zero-bias value $dV/dI(I=0)$ is stable in parallel magnetic field, while the width of the central 
region is gradually decreasing, see Fig.~\ref{mag} (b). Subgap ABS peaks monotonicly come to nearly-zero position, they 
are  coalescing together at approximately 2~T, see Fig.~\ref{colorplot} (c). The stability of the  zero-bias level 
$dV/dI(I=0)$ below 2~T is also demonstrated by the $dV/dI(I=0)$ magnetic field scan  in Fig.~\ref{colorplot} (d) for 
parallel magnetic field.

Thus, evolution of $dV/dI(I)$ curves is drastically different for two field orientations. Qualitatively similar behavior
can be observed for In-CoSi-In junctions, as it is depicted in  Fig.~\ref{mag} (c). The zero-bias value $dV/dI(I=0)$ is
very sensitive to normal magnetic field 48~mT, while the dependence is weak for the in-plane magnetic field of the same 
value. $dV/dI(I=0)$ magnetic field scans support this similarity in the insets to Fig.~\ref{colorplot} (b) and (d), 
despite much smaller (about 40~mT) onset field for the In-CoSi-In junctions.

Due to the well-defined superconducting gap in the experimental data, the observed $dV/dI(V)$ features should be connected with spectrum specifics and they can not be ascribed to current-induced effects. A rough estimation of heating in the contact area may be done by using the bias voltage (mV) by a factor of 3.2 K/mV~\cite{heating}. For the observed $dV/dI(V)$ features $V$ = 0.2~mV gives heating much less than Nb $T_c$, so we can not attribute the ZBA to the heating effects.

As a result, we observe coalescence of $dV/dI$ subgap ABS peaks at nearly-zero bias  and perfect stability of the 
zero-bias $dV/dI(I=0)$ value only for the parallel to the CoSi flake surface magnetic fields. These effects are qualitatively
similar for In or Nb superconducting leads, so they reflect behavior of proximized CoSi surface.

\begin{center}
\begin{figure}
\includegraphics[width=\columnwidth]{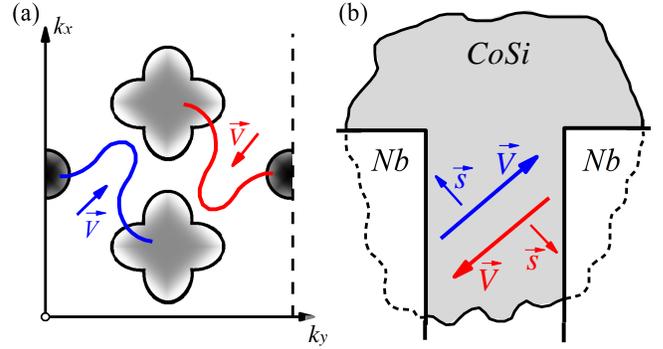}
\caption{(Color online) (a) Sketch of Fermi arcs (color) and bulk spectrum projection (black) on (001) surface Brillouin zone, in accordance with ARPES data in Ref.~\onlinecite{cosi1}. In a chiral topological semimetal there is only one pair of chiral 
nodes of opposite Chern numbers with large separation in momentum space. This leads  to extremely long surface Fermi 
arcs~\cite{long}, these arcs  define a specific transmission direction of carriers in topological surface states.
(b) Sketch of the surface transport between two Nb leads. Surface transport is effectively one-dimensional due to the specific transmission direction of carriers in topological surface states. An inhomogeneous potential is formed at the surface of CoSi by two interfaces with the Nb leads, providing a 
platform for confined ABSs. The spin-orbit coupling-induced splitting near the Fermi surface in CoSi should be considered to be sufficiently strong as far
as the proximity-induced superconductivity is concerned. Thus, ABSs coalescence can be expected due to the joined effect of spin-orbit coupling  and Zeeman interaction, similarly to the proximized semiconductor nanowires~\cite{sarma}. $S$ indicates the in-plane spin polarization of the Fermi arc surface states in CoSi~\cite{Burkovetal2018}, which is responsible for the pronounced dependence of the subgap ABS $dV/dI$ peaks on the magnetic field direction in Figs.~\ref{mag}, and~\ref{colorplot}  (see the main text).
 }
\label{Arcs}
\end{figure}
\end{center}

ABS behavior in Fig.~\ref{colorplot} (c)  strongly resembles the one taking place for  proximized semiconductor nanowires, whose conductance spectra 
have been extensively studied both experimentally \cite{mourik2012,das2012,deng2012,chang2015,albrecht2016,mzm2,%
chen2017,gul2017,zhang2017,gul2018,grivnin2019,chen2019} and theoretically (see, e.g. Ref~\onlinecite{sarma} and 
references therein).  In the absence of an intrinsic Zeeman splitting as well as an external magnetic 
field, non-topological ABSs usually appear as two symmetric subgap peaks~\cite{mzm2,ABS}. However, when the Zeeman field 
increases in the presence of a pronounced SOC, the finite-energy Andreev states can coalesce together and form 
near-zero-energy midgap states~\cite{mzm2}.  The theory predicts that the topological phase transition, where almost-zero-energy Andreev bound 
states transform into the Majorana modes, takes place at the Zeeman field well exceeding the coalescence point. It is 
difficult to experimentally distinguish between the topological and non-topological scenarios under such conditions, 
since they both result in the zero-bias peak \cite{sarma}. The main qualitative statement of this theoretical model can 
be also applied to two-dimensional systems~\cite{supercond,flensberg,sato,woods2019,vuik2019}. 

Similar physical mechanism can be responsible for ABS coalescence together at the surface of chiral semimetal CoSi in 
proximity to two Nb superconducting leads:  

(i) Although surface states are generally two-dimensional, a preferable direction is defined by Fermi arcs on a 
particular crystal surface~\cite{nbwte}. This should be especially significant for chiral semimetals with long Fermi 
arcs~\cite{long,cosi1,cosi2}, see Fig.~\ref{Arcs} (a).

(ii) An inhomogeneous potential is formed at the surface of CoSi by two interfaces with the Nb leads, providing a 
platform for confined ABSs. Superconducting correlations can be efficiently induced in topological surface states even 
for several-micrometer-long junctions~\cite{inwte1,inwte2,incosns,ali,liao}.

(iii) As it is well known, the condition of a pronounced SOC is satisfied for the standard topological semimetal surface
states~\cite{armitage}. The relative strength of SOC in chiral topological semimetal CoSi is of the 
order of millielectronvolts due to the weak SOC on the Co $3d$ and Si $3p$ orbitals \cite{Burkovetal2018}. Since such an
energy scale of the SOC-induced band splitting is much less than the characteristic interband separation (see, for 
example, Figs. 3A and S4 in Ref.~\cite{Xuetal2020}), the effects of SOC is characterized as small in forming the CoSi 
band structure and the topological surface states.

This does not concern, however, the influence of SOC on the superconducting effects, where significantly smaller 
energetic scale near the Fermi surface comes to the fore. The SOC-induced splitting near the Brillouin zone center (the 
$\Gamma$ point), estimated experimentally as $\approx 18.1$~meV \cite{Xuetal2020}, is large compared with the 
superconducting scale and encompasses the Fermi level. Since the splitting is known to substantially modify the 
topological structure of the multifold crossings, changing the number of bands that meet at the high-symmetry point 
\cite{Changetal2017,zhang}, the SOC near the Fermi surface in CoSi should be considered to be sufficiently strong as far
as the proximity-induced superconductivity is concerned.

One of the striking features observed in this paper is a pronounced dependence of the subgap ABS $dV/dI$ peaks on the 
magnetic field direction in Figs.~\ref{mag}, and~\ref{colorplot}. This anisotropy should be associated with the spin 
polarization of the Fermi arc surface states in CoSi. 

A spin polarization of the Fermi arcs is known to take place in a number of topological Weyl semimetals due to a strong
spin-momentum locking. Thus the spin polarization of the arcs, that has been discovered in TaAs, lies completely in the 
plane of the (001) surface and reaches $80\%$  \cite{Xuetal2016}. Recent theoretical studies of chiral topological
semimetal CoSi have shown that, in disregarding the effects of SOC, the Fermi arcs are spin degenerate \cite{zhang}. 
However, the spin-orbit interaction lifts the spin degeneracy of the surface states leading to their in-plane spin 
polarization on the (001) surface, with strongly correlated and predominantly antiparallel spin textures in the 
neighboring Fermi arcs \cite{Burkovetal2018}. 

This fully supports our interpretation of the experimental results in Fig.~\ref{colorplot} (a). The in-plane spin 
polarization of the Fermi arc surface states in CoSi allows the Zeeman interaction only with the in-plane components of 
the magnetic field and not with its normal to the surface component. The finite-energy ABS peaks coalesce together due 
to the interaction of the spin textures of individual arcs with increasing in-plane Zeeman field, in the presence of 
SOC.  	

While the topological transition to the state with Majorana fermions is generally expected to occur at a Zeeman field 
well exceeding the coalescence point \cite{sarma} (about 2~T in Fig.~\ref{colorplot} (c)), the zero-bias $dV/dI(I=0)$ 
level stability is destroyed above the same 2~T field in Fig.~\ref{colorplot} (d), probably due to the close value of 
the Nb film critical field. Therefore, the topological transition point seems to be unreachable in our samples, although
the surface Majorana fermions are generally allowed in superconducting topological chiral semimetal 
CoSi~\cite{supercond} and we cannot experimentally distinguish between the topological and non-topological scenarios in
Fig.~\ref{colorplot} (c). 

Although the direct and inverse magnetoelectric Edelstein effects are known to be generally present in 
noncentrosymmetric materials and superconductors 
\cite{LevitovNazarovEliashberg1985,LevitovNazarovEliashberg1985_2,AronovLyandaGeller1989,Edelstein1990,Edelstein1995,%
KatoetalAwschalom2004,Edelstein2005,LuYip2008,ShenVignaleRaimondi2014,Yip2014,Smidmanetal2017,HeLaw2020}, we have 
detected no conclusive evidence of them in our study.


The Andreev transport through a region of proximity-induced superconductivity in topological chiral semimetal CoSi has 
demonstrated the coalescence of ABS peaks to nearly-zero bias position and the stability of the zero-bias $dV/dI$ 
resistance value with increasing parallel to the flake surface magnetic field. Normal to the surface field only
monotonicly suppresses the peaks. We associate the striking magnetic field anisotropy with the Zeeman interaction of the 
in-plane polarized spin texture of the Fermi arcs in CoSi. The Andreev states behavior is interpreted as the joined 
effect of the strong SOC and the Zeeman interaction, known for proximized semiconductor nanowires. The 
topological transition point to the state with Majorana fermions seems to be unreachable in our samples.


We wish to thank V.T. Dolgopolov for fruitful discussions, and S.S~Khasanov for X-ray sample characterization.
We gratefully acknowledge financial support partially by the RFBR  (project No.~19-02-00203), and RF
State task.


\begin{thebibliography}{99}


\bibitem{bernevig} B. Bradlyn, J. Cano, Z. Wang, M. G. Vergniory, C. Felser, R. J. Cava, and B. A. Bernevig, 
Science 353, aaf5037 (2016)
\bibitem{zhang} Peizhe Tang, Quan Zhou, and Shou-Cheng Zhang, Phys. Rev. Lett. 119, 206402 (2017)
\bibitem{armitage} As a recent review see N.P.~Armitage, E.J.~Mele, and A.~Vishwanath,  
Rev. Mod. Phys.  90, 015001 (2018).

\bibitem{magnet} S. Zhong, J.E. Moore, and I. Souza,  Phys. Rev. Lett. 116, 077201 (2016).
\bibitem{lattice} P. Rinkel, P. L. S.  Lopes, and I. Garate,  Phys. Rev. Lett. 119, 107401 (2017).
\bibitem{photo} F. de Juan, A.G. Grushin, T. Morimoto, and J.E. Moore,  Nat. Commun. 8, 15995 (2017).
\bibitem{long} N. B. Schr\"oter, D. Pei, M. G. Vergniory, Y. Sun, K. Manna, F. de Juan, J. A. Krieger, V. S\"uss, 
M. Schmidt, P. Dudin, B. Bradlyn, T. K. Kim, Th. Schmitt, C. Cacho, C. Felser, V. N. Strocov and Y. Chen, Nature Physics 15, 759 (2019).



\bibitem{cosi1} Zhicheng Rao, Hang Li, Tiantian Zhang, Shangjie Tian, Chenghe Li, Binbin Fu, Cenyao Tang, Le Wang, 
Zhilin Li, Wenhui Fan, Jiajun Li, Yaobo Huang, Zhehong Liu, Youwen Long, Chen Fang, Hongming Weng, Youguo Shi, 
Hechang Lei, Yujie Sun, Tian Qian and Hong Ding, Nature 567, 496 (2019).
\bibitem{cosi2} Daichi Takane, Zhiwei Wang, Seigo Souma, Kosuke Nakayama, Takechika Nakamura,  Hikaru Oinuma, Yuki 
Nakata,  Hideaki Iwasawa,  Cephise Cacho,  Timur Kim,  Koji Horiba,  Hiroshi Kumigashira, Takashi Takahashi, Yoichi 
Ando,  and Takafumi Sato, Phys. Rev. Lett. 122, 076402 (2019).
\bibitem{maxChern} Niels B. M. Schr{\"o}ter, Samuel Stolz, Kaustuv Manna, Fernando de Juan, Maia G. Vergniory, 
Jonas A. Krieger, Ding Pei, Thorsten Schmitt, Pavel Dudin, Timur K. Kim, Cephise Cacho, Barry Bradlyn, Horst Borrmann, 
Marcus Schmidt, Roland Widmer, Vladimir N. Strocov, Claudia Felser, Science 369, 179 (2020).




\bibitem{review1} C. W. J. Beenakker, Annu. Rev. Con. Mat. Phys. 4, 113 (2013).
\bibitem{review2}  J. Alicea, Rep. Prog. Phys. 75, 076501 (2012).
\bibitem{LiXu2019} Yupeng Li, Zhu-An Xu, Adv. Quant. Technol. 2, 1800112 (2019).
\bibitem{volovik} G.E. Volovik, JETP Letters 107, 516 (2018)

\bibitem{zhang1} S. Murakami, N. Nagaosa, S.-C. Zhang, Phys. Rev. Lett. 93, 156804 (2004).
\bibitem{kane} C. L. Kane, E. J. Mele, Phys. Rev. Lett. 95, 146802 (2005).
\bibitem{zhang2} B. A. Bernevig, S.-C. Zhang, Phys. Rev. Lett. 96, 106802 (2006).
\bibitem{Fu} L. Fu and C. L. Kane,  Phys. Rev. Lett. 100, 96407 (2008).

\bibitem{spec} Wei Chen,  Liang Jiang,  R. Shen,  L. Sheng,  B. G. Wang,  D. Y. Xing, EPL 103, 27006 (2013)
\bibitem{been1} C. W. J. Beenakker, Physical Review Letters 97 (2006).
\bibitem{been2} C. W. J. Beenakker, Reviews of Modern Physics 80, 1337 (2008).

\bibitem{Franz2016} A. Chen and M. Franz, Phys. Rev. B 93,  201105 (2016).
\bibitem{arcsc} Z. Faraei and S. A. Jafari, Phys. Rev. B 100, 035447 (2019).
\bibitem{dutta2020} P. Dutta, F. Parhizgar, and A.~M. Black-Schaffer, Phys. Rev. B 101, 064514 (2020).

\bibitem{inwte1} O.O. Shvetsov, A. Kononov, A.V. Timonina, N.N. Kolesnikov, E.V. Deviatov, 
JETP Letters, 107, 774 (2018).
\bibitem{inwte2} O.O. Shvetsov, A. Kononov, A.V. Timonina, N.N. Kolesnikov, E.V. Deviatov, EPL, 124, 47003 (2018).
\bibitem{incosns} O. O. Shvetsov, V. D. Esin, Yu. S. Barash, A. V. Timonina, N. N. Kolesnikov, and E. V. Deviatov, 
Phys. Rev. B 101, 035304 (2020)
\bibitem{nbwte} A. Kononov, O.O. Shvetsov, S.V. Egorov, A.V. Timonina, N.N. Kolesnikov and E.V. Deviatov, 
EPL, 122, 27004 (2018)


\bibitem{supercond} Yingyi Huang and Shao-Kai Jian, arXiv:2009.04654

\bibitem{sarma} Chun-Xiao Liu, Jay D. Sau, Tudor D. Stanescu, S. Das Sarma, Phys. Rev. B 96, 075161 (2017).

\bibitem{Burkovetal2018} D. A. Pshenay-Severin, Y. V. Ivanov, A. A. Burkov, A. T. Burkov, 
Journal of Physics: Condensed Matter 30, 135501 (2018).


\bibitem{cdas} O. O. Shvetsov, V. D. Esin, A. V. Timonina, N. N. Kolesnikov, and E. V. Deviatov
	Phys. Rev. B 99, 125305 (2019)
\bibitem{magres}  D. S. Wu, Z. Y. Mi, Y. J. Li, W. Wu, P. L. Li, Y. T. Song, G. T. Liu, G. Li, J. L. Luo,  Chinese Phys. Lett. 36 077102 (2019)
\bibitem{indium} A. M. Toxen Phys. Rev. 123, 442 (1961).
\bibitem{andreev} A. F. Andreev, Soviet Physics JETP {\bf 19}, 1228 (1964).
\bibitem{tinkham} M. Tinkham, Introduction to Superconductivity (2d ed., McGraw–Hill, New York, 1996).
\bibitem{nbcosns} O. O. Shvetsov, Yu. S. Barash, S. V. Egorov, A. V. Timonina, N. N. Kolesnikov, and E. V. Deviatov, 	EPL, 132, 67002 (2020)


\bibitem{ABS} Luca Banszerus, Florian Libisch, Andrea Ceruti, Stefan Blien, Kenji Watanabe, Takashi Taniguchi, 
Andreas K. Hüttel, Bernd Beschoten, Fabian Hassler, Christoph Stampfer, arXiv:2011.11471

\bibitem{heslinga} D.R.~Heslinga, S.E.~Shafranjuk, H.~van~Kempen, and T.M.~Klapwijk, 
Phys. Rev. B {\bf 49}, 10484 (1994).




\bibitem{klapwijk17}  J. Wiedenmann, E. Liebhaber, J.s K\"ubert, E. Bocquillon, Ch. Ames, H. Buhmann, T.M. Klapwijk, 
L.W. Molenkamp,  Phys. Rev. B 96, 165302 (2017).
\bibitem{ingasb} A. Kononov, V.A. Kostarev, B.R. Semyagin, V.V. Preobrazhenskii, M.A. Putyato, E.A. Emelyanov, 
E.V. Deviatov, Physical Review B 96, 245304 (2017)

\bibitem{osbite} O.O. Shvetsov, V.A. Kostarev, A. Kononov,  V.A. Golyashov, K.A. Kokh, O.E. Tereshchenko, E.V. Deviatov,
EPL 119, 57009 (2017); DOI: 10.1209/0295-5075/119/57009.
\bibitem{tomasch1} W. J. Tomasch, Phys. Rev. Lett. 16, 16 (1966).
\bibitem{tomasch2}  W. L. McMillan, P. W. Anderson, Phys. Rev. Lett. 16, 85 (1966).
\bibitem{mcmillan1} J. M. Rowell, W. L. McMillan, Phys. Rev. Lett. 16, 453 (1966).
\bibitem{mcmillan2} J. M. Rowell,  Phys. Rev. Lett. 30, 167 (1973).
\bibitem{kuz1} T. E. Kuzmicheva, S. A. Kuzmichev, A. V. Sadakov, A. V. Muratov, A. S. Usoltsev, V. P. Martovitsky, A. R. Shipilov, D. A. Chareev, E. S. Mitrofanova and V. M. Pudalov, JETP Letters 104, 858 (2016) 
\bibitem{kuz2} T. E. Kuzmicheva, S. A. Kuzmichev and N. D. Zhigadlo,  JETP Letters 112, 497 (2020)


\bibitem{mzm2} M.~T. Deng, S. Vaitiekėnas, E.~B. Hansen, J. Danon, M. Leijnse, K. Flensberg, J. Nyg\'ard, 
P. Krogstrup, C.~M. Marcus, Science 354, 1557 (2016).
\bibitem{adroguer} P.~Adroguer, C.~Grenier, D.~Carpentier, J.~Cayssol, P.~Degiovanni, and E.~Orignac, Phys. Rev. B {\bf 82}, 081303(R), (2010).

\bibitem{heating} Y.~G.Naidyuky, I.~K. Yanson, J. Phys. Condens. Matter, 10:8905–38 (1998).

\bibitem{mourik2012} V. Mourik, K. Zuo, S.~M. Frolov, S.~R. Plissard, E.~P.~A.~M. Bakkers, L.~P. Kouwenhoven, 
Science 336, 1003 (2012).
\bibitem{das2012}  A. Das, Y. Ronen, Y. Most, Y. Oreg, M. Heiblum, and H. Shtrikman, Nat. Phys. 8, 887 (2012).
\bibitem{deng2012} M.~T. Deng, C.~L. Yu, G.~Y. Huang, M. Larsson, P. Caroff, and H.~Q. Xu, 
Nano Lett. 12, 6414 (2012).
\bibitem{chang2015} W. Chang, S.~M. Albrecht, T.~S. Jespersen, F. Kuemmeth, P. Krogstrup, J. Nyg\'ard,
and C.~M. Marcus, Nat. Nanotech. 10,232 (2015).
\bibitem{albrecht2016} S.~M. Albrecht, A.~P. Higginbotham, M. Madsen, F. Kuemmeth, T.~S. Jespersen, 
J. Nyg\'ard,  P. Krogstrup, and C.~M. Marcus, Nature 531, 206 (2016).

\bibitem{chen2017} J. Chen, P. Yu, J. Stenger, M. Hocevar, D. Car, S.~R. Plissard, 
E.~P.~A.~M. Bakkers, T.~D. Stanescu, S.~M. Frolov, Sci. Adv. 3, e1701476 (2017).
\bibitem{gul2017}  \"{O}. G\"{u}l, H. Zhang, F.~K. de Vries, J. van Veen, K. Zuo, V. Mourik, S. Conesa-Boj,
M.~P. Novak, D.~J. van Woerkom, M. Quinetro-Perez, M.~C. Cassidy, A. Geresdi, S. Koelling, D. Car,
S.~R. Plissard, E.~P.~A.~M. Bakkers, and L.~P. Kouwenhoven,  Nano Lett. 17, 2690 (2017).
\bibitem{zhang2017} H. Zhang, \"{O}. G\"{u}l, S. Conesa-Boj, M.~P. Novak, M. Wimmer, K. Zuo,
V. Mourik, F.~K. de Vries, J. van Veen, M.~W.~A. de~Moor, J.~D.~S. Bommer, D.~J. van Woerkom, 
D. Car, S.~R. Plissard, E.~P.~A.~M. Bakkers, M. Quinetro-Perez, M.~C. Cassidy, S. Koelling,
S. Goswami, K. Watanabe, T. Taniguchi, and L.~P. Kouwenhoven, Nat. Comm. 8, 16025 (2017).
\bibitem{gul2018}  \"{O}. G\"{u}l,  H. Zhang, J.~D.~S. Bommer, M.~W.~A. de~Moor, D. Car, 
S.~R. Plissard, E.~P.~A.~M. Bakkers, A. Geresdi, K. Watanabe, T. Taniguchi, and L.~P. Kouwenhoven,
Nat. Nanotech. 13, 192 (2018).
\bibitem{grivnin2019} A. Grivnin, E. Bor, M. Heiblum, Y. Oreg, and H. Shtrikman, Nat. Com. 10, 1940 (2019).
\bibitem{chen2019}  J. Chen, B.~D. Woods, P. Yu, M. Hocevar, D. Car, S.~R. Plissard,
E.~P.~A.~M. Bakkers, T.~D. Stanescu, and S.~M. Frolov, Phys. Rev. Lett. 123, 107703 (2019).



\bibitem{flensberg} M. Leijnse and K. Flensberg, Semicond. Sci. Technol. 27, 124003 (2012).
\bibitem{sato} M. Sato, S. Fujimoto, J. Phys. Soc. Jpn. 85, 072001 (2016).
\bibitem{woods2019} B.~D. Woods, J. Chen, S.~M. Frolov, and T.~D. Stanescu, Phys. Rev. B 100, 125407 (2019).
\bibitem{vuik2019} A. Vuik, B. Nijholt, A.~R. Akhmerov and M. Wimmer, SciPost Phys. 7, 061 (2019).




\bibitem{ali} Yaojia Wang, Shuoying Yang, Pranava K. Sivakumar, Brenden R. Ortiz, Samuel M.L. Teicher, Heng Wu, 
Abhay K. Srivastava, Chirag Garg, Defa Liu, Stuart S. P. Parkin, Eric S. Toberer, Tyrel McQueen, Stephen D. Wilson, 
Mazhar N. Ali, 	arXiv:2012.05898. 
\bibitem{liao} Cai-Zhen Li, An-Qi Wang, Chuan Li, Wen-Zhuang Zheng, Alexander Brinkman, Da-Peng Yu, and Zhi-Min Liao, 
Phys. Rev. Lett. 124, 156601 (2020).


\bibitem{Xuetal2020} Bing Xu, Zhenyao Fang, Miguel-{\'A}ngel S{\'a}nchez-Mart{\'\i}nez, Jorn W. F. Venderbos, 
Zhuoliang Ni, Tian Qiu, Kaustuv Manna, Kefeng Wang, Johnpierre Paglione, Christian Bernhard, Claudia Felser, 
Eugene J. Mele, Adolfo G. Grushin, Andrew M. Rappe, Liang Wu, PNAS 117, 27104 (2020).


\bibitem{Changetal2017} Guoqing Chang, Su-Yang Xu, Benjamin J. Wieder, Daniel S. Sanchez, Shin-Ming Huang, 
Ilya Belopolski, Tay-Rong Chang, Songtian Zhang, Arun Bansil, Hsin Lin, M. Zahid Hasan, PRL 119, 206401 (2017).



\bibitem{Xuetal2016} Su-Yang Xu, Ilya Belopolski, Daniel S. Sanchez, Madhab Neupane, Guoqing Chang, Koichiro Yaji, 
Zhujun Yuan, Chenglong Zhang, Kenta Kuroda, Guang Bian, Cheng Guo, Hong Lu, Tay-Rong Chang, Nasser Alidoust, Hao Zheng, 
Chi-Cheng Lee, Shin-Ming Huang, Chuang-Han Hsu, Horng-Tay Jeng, Arun Bansil, Titus Neupert, Fumio Komori, Takeshi Kondo, 
Shik Shin, Hsin Lin, Shuang Jia, M. Zahid Hasan, Phys. Rev. Lett. 116, 096801 (2016).

\bibitem{LevitovNazarovEliashberg1985} L. S. Levitov, Yu. V. Nazarov, G. M. Eliashberg, JETP Lett. 41, 445 (1985).
\bibitem{LevitovNazarovEliashberg1985_2} L. S. Levitov, Yu. V. Nazarov, G. M. Eliashberg, JETP 61, 133 (1985).
\bibitem{AronovLyandaGeller1989} A. G. Aronov and Y. B. Lyanda-Geller, JETP Lett. 50, 431 (1989).
\bibitem{Edelstein1990} V. M. Edelstein, Solid State Commun. 73, 233 (1990).
\bibitem{Edelstein1995} V. M. Edelstein, PRL 75, 2004 (1995).
\bibitem{KatoetalAwschalom2004} Y. K. Kato, R. C. Myers, A. C. Gossard, D. D. Awschalom, PRL 93, 176601 (2004).
\bibitem{Edelstein2005} V. M. Edelstein, Physical Review B 72, 172501 (2005).
\bibitem{LuYip2008} C-K. Lu, S. Yip, Physical Review B 77, 054515 (2008).
\bibitem{Yip2014} S. Yip Annu. Rev. Condens. Matter Phys. 5, 15 (2014).
\bibitem{ShenVignaleRaimondi2014} Ka Shen, G. Vignale, R. Raimondi, PRL 112, 096601 (2014).
\bibitem{Smidmanetal2017} M. Smidman, M. B. Salamon, H. Q. Yuan, D. F. Agterberg Rep. Prog. Phys. 80, 036501 (2017).
\bibitem{HeLaw2020} Wen-Yu He, K. T. Law, Phys. Rev. Research 2, 012073 (2020).



\end{thebibliography}
\end{document}